\documentclass{emulateapj}

\usepackage{times}
\usepackage[]{graphicx,epsfig,rotating}

\begin{document}

\bibliographystyle{apj}
\title{On the Fraction of Barred Spiral Galaxies}
\author{\medskip Preethi B Nair\footnote{Present address: INAF - Astronomical Observatory of Bologna, Via Ranzani 1, 40127 Bologna, ITALY}   \&\ Roberto G Abraham}
\affil{
   Department of Astronomy \& Astrophysics,
   University of Toronto, 50 St. George Street,
   Toronto, ON, M5S~3H4.
}
\slugcomment{Submitted June 16, 2009; Accepted by ApJ Letters April 6, 2010}

\email{preethi.nair@oabo.inaf.it, abraham@astro.utoronto.ca}


\begin{abstract}
We investigate the
stellar masses of strongly barred spiral galaxies. Our analysis is based on 
a sample of  $\thicksim$14000 visually-classified nearby galaxies given
in \cite{Nair:2010p22790}. The fraction of barred spiral galaxies is found to be a strong
function of stellar mass and star formation history, with a minimum near
the characteristic mass at which bimodality is seen in the stellar populations of galaxies.
We also find bar fractions are very sensitive to the central concentration of galaxies below the transition mass but not above it.
This suggests  that whatever process is causing the creation of the red and blue sequences is either influencing, 
or being influenced by, structural changes which manifest themselves in the absence of bars.  
As a consequence of strong bar fractions being sensitive to the mass range probed, our analysis helps resolve discrepant results on the reported evolution of bar fractions with redshift.

\end{abstract}

\section{INTRODUCTION}
\label{sec:introduction}
Understanding the role of bars in galaxy formation is central to understanding the evolution of galaxies. 
Bars are important structures that help to redistribute angular momentum between baryonic and dark  matter components in disk galaxies \citep{Weinberg:1985p8431,Debattista:2000p8516,Athanassoula:2002p7988} thereby driving their secular and dynamic evolution \citep{Kormendy:2004p9826}. Bars are thought to drive spiral arms \citep{Lindblad:1960p8377,Toomre:1969p8347,Sanders:1976p8346} and ring structures \citep{Schwarz:1981p8243,Buta:1996p2930,MartinezValpuesta:2006p251}.
They transport gas/matter  to the centers of galaxies \citep{Knapen:1995p8521,Hunt:1999p8545} and help to build bulges \citep{Laurikainen:2007p7254} and possibly trigger AGN activity \citep{Laine:2002p7546,Knapen:2000p266,Laurikainen:2004p7257}.  

The importance of bars in galaxy evolution has motivated a number of 
recent studies.
In a comprehensive investigation of the fraction of barred spirals as a function of cosmic epoch, 
\cite{Sheth:2008p1183} (hereafter SE08) used a sample of  $\thicksim 2000$ galaxies from the Hubble Space Telescope
COSMOS survey \citep{Scoville:2007p1332} to show that the bar fraction decreases with redshift, as claimed by \cite{Abraham:1999p450} and \cite{vandenBergh:2002p9582} (but see \cite{Elmegreen:2004p1773} and \cite{Jogee:2004p16823}). 
In addition, SE08 also find that the bar fraction of spiral galaxies is a strong function of stellar mass, color and bulge prominence such that more massive, redder, concentrated galaxies have a larger bar fraction than less massive, bluer, diskier galaxies. 
These observations are rather different from those presented by
\cite{Barazza:2008p82} (hereafter BJ08), who examined a sample of 
$\thicksim 2000$ galaxies from the Sloan Digital Sky Survey in the redshift range between 0.01 and 0.03. In sharp contrast with SE08, 
these authors
claim that bar fractions increase with decreasing mass and bluer colors (corresponding to late type galaxies). 

Can the observations of SE08 and BJ08 be reconciled? One possibility is that the different claims represent an evolutionary
effect. This seems unlikely because the trends reported by SE08 apply equally well in the lowest redshift bin of that 
investigation.
A more promising explanation lies in the fact that,
for obvious reasons,
magnitude-limited surveys of high-redshift galaxies will always tend to probe more massive galaxies than those
probed by low-redshift surveys.
Therefore the stellar mass ranges spanned by the galaxies
in SE08 and BJ08 only partially overlap. SE08 is not sensitive to galaxies with stellar masses M $<$$10^{10.2}$$M_{\odot}$ (thus missing
many dwarf systems) while BJ08 is not sensitive to galaxies with stellar masses M$>$$10^{10.5}$$M_{\odot}$
(thus missing the most luminous and massive objects in the local universe). It therefore seems quite conceivable that the apparent discrepancy in these studies is due to differences in the stellar mass ranges being probed.

To test this hypothesis, and to learn more about the nature of barred spiral galaxies at a range of
masses, we use the sample of 14034 visually classified galaxies from \cite{Nair:2010p22790}(hereafter Paper I).
The reader is referred to Paper I for details but in summary all spectroscopically targeted galaxies from the 
SDSS DR4 \citep{Stoughton:2002p1611,York:2000p3192}, with an extinction corrected g-band 
magnitude g$<$16 at redshifts between 0.01 and 0.1, were visually classified by one of the 
authors (PN) using the Carnegie Atlas of Galaxies \citep{Sandage:1994p4888} as a visual training set. 
Comparisons of our classification with the Third Reference Catalog
of Bright Galaxies \citep{deVaucouleurs:1991p4597}(RC3) for the $\thicksim$1700 objects in common showed excellent agreement with a mean deviation of 1.2 T-Types)\footnote{An easy way to remember the numeric equivalent for a T-Type is to note that all major classes are odd numbers (e.g. Sa galaxies have T-Type 1, and Sb galaxies have T-Type 3), while the finer separation are even numbers (e.g. Sab has T-Type 2)}.
Bars in our sample were visually identified and are equivalent to `strong bars' in the RC3 catalog (as noted in Paper I). In order to simplify our investigation, we restrict ourselves to disk galaxies (which we define to be S0 galaxies and later with axial ratios b/a$>$0.4). The bar fraction for this sample is $\thicksim$ 30\%, with 2312 barred galaxies. We use the stellar masses derived by \cite{Kauffmann:2003p97} for the following analysis.

\section{Stellar Mass and Color Distributions}

\begin{figure}[t!]
\unitlength1cm
\begin{minipage}[t]{4.0cm}
\rotatebox{0}{\resizebox{4cm}{4cm}{\includegraphics{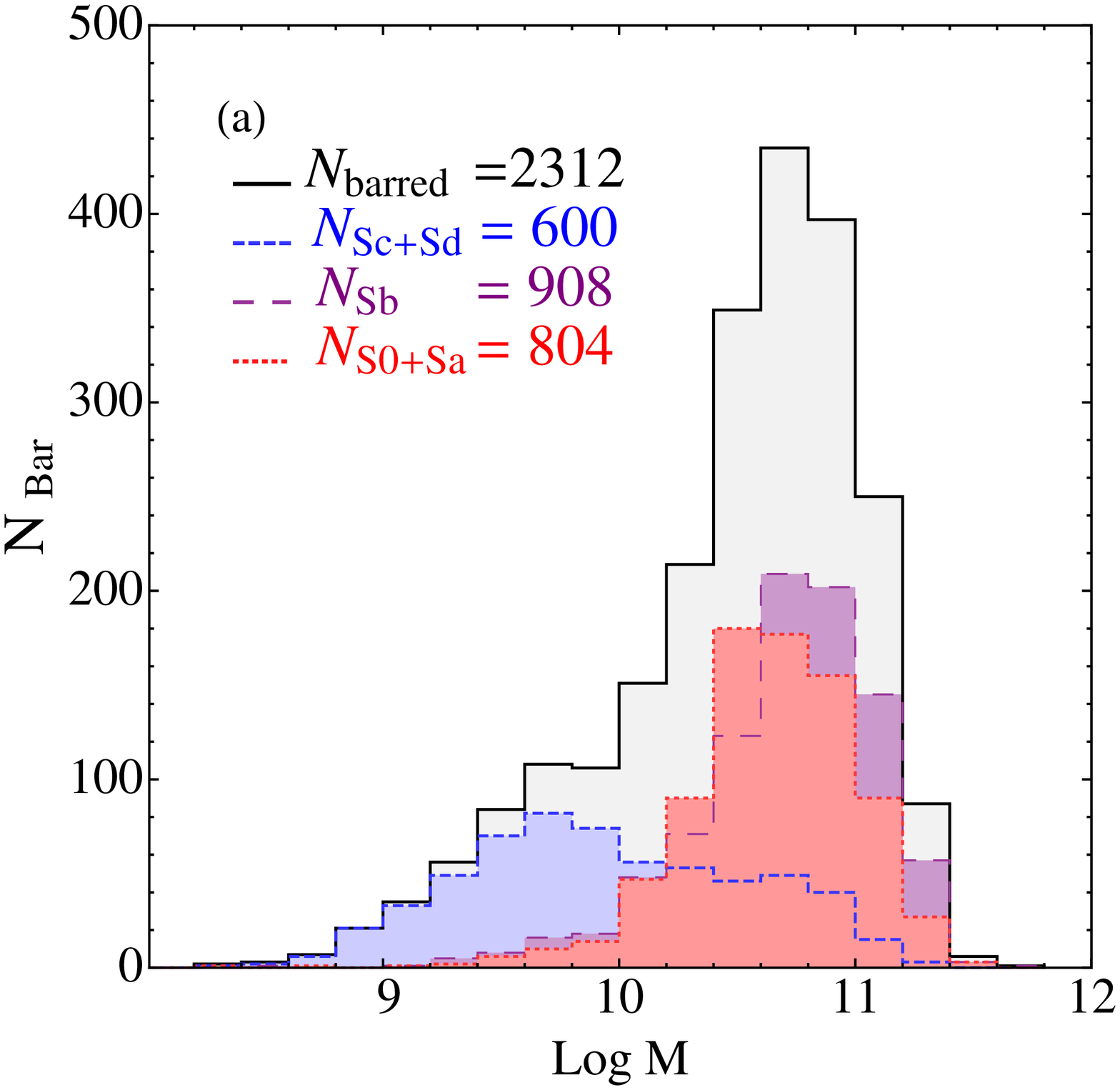}}}
\rotatebox{0}{\resizebox{4cm}{4cm}{\includegraphics{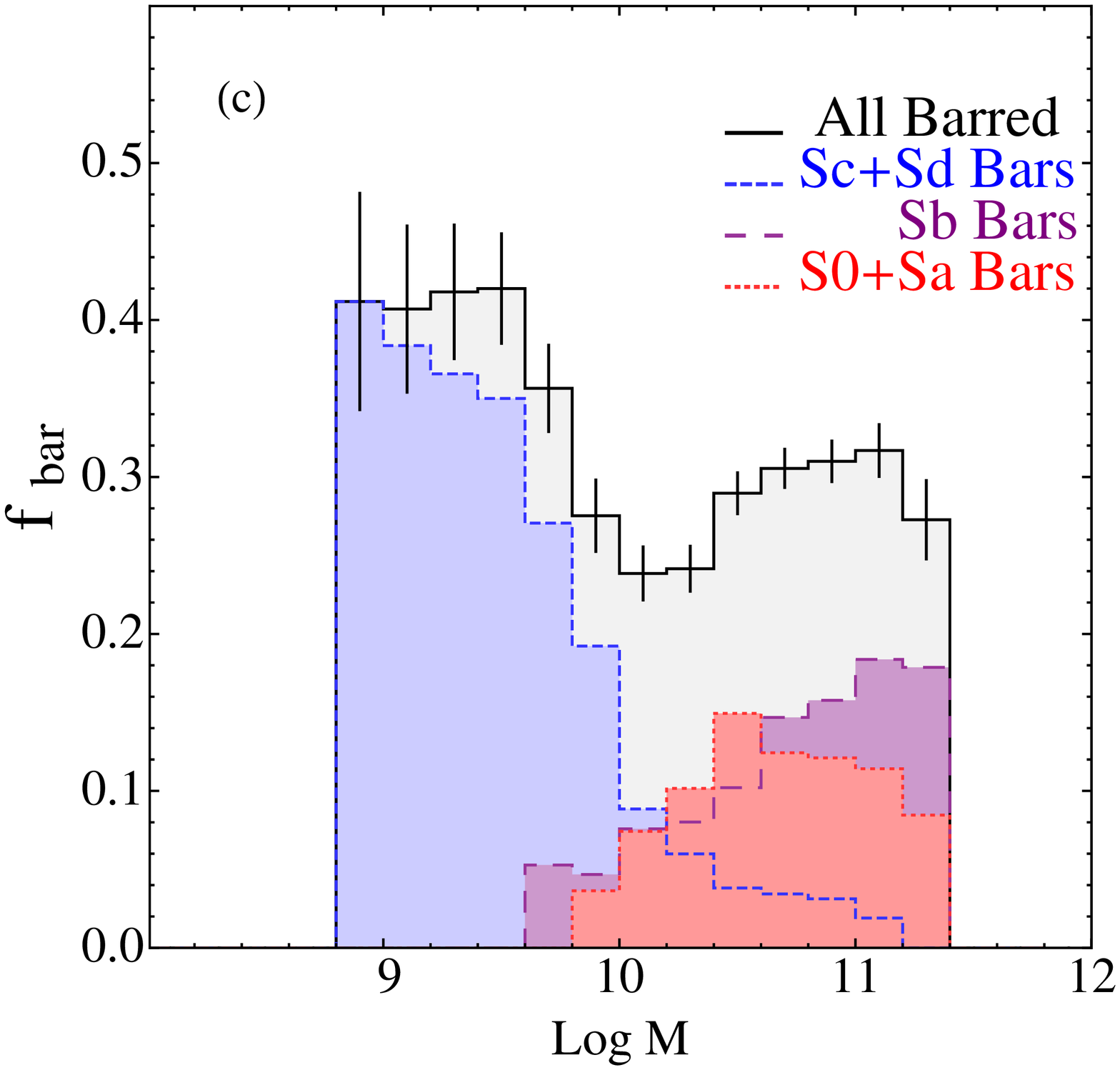}}}
\end{minipage}
\begin{minipage}[t]{4.0cm}
\rotatebox{0}{\resizebox{4cm}{4cm}{\includegraphics{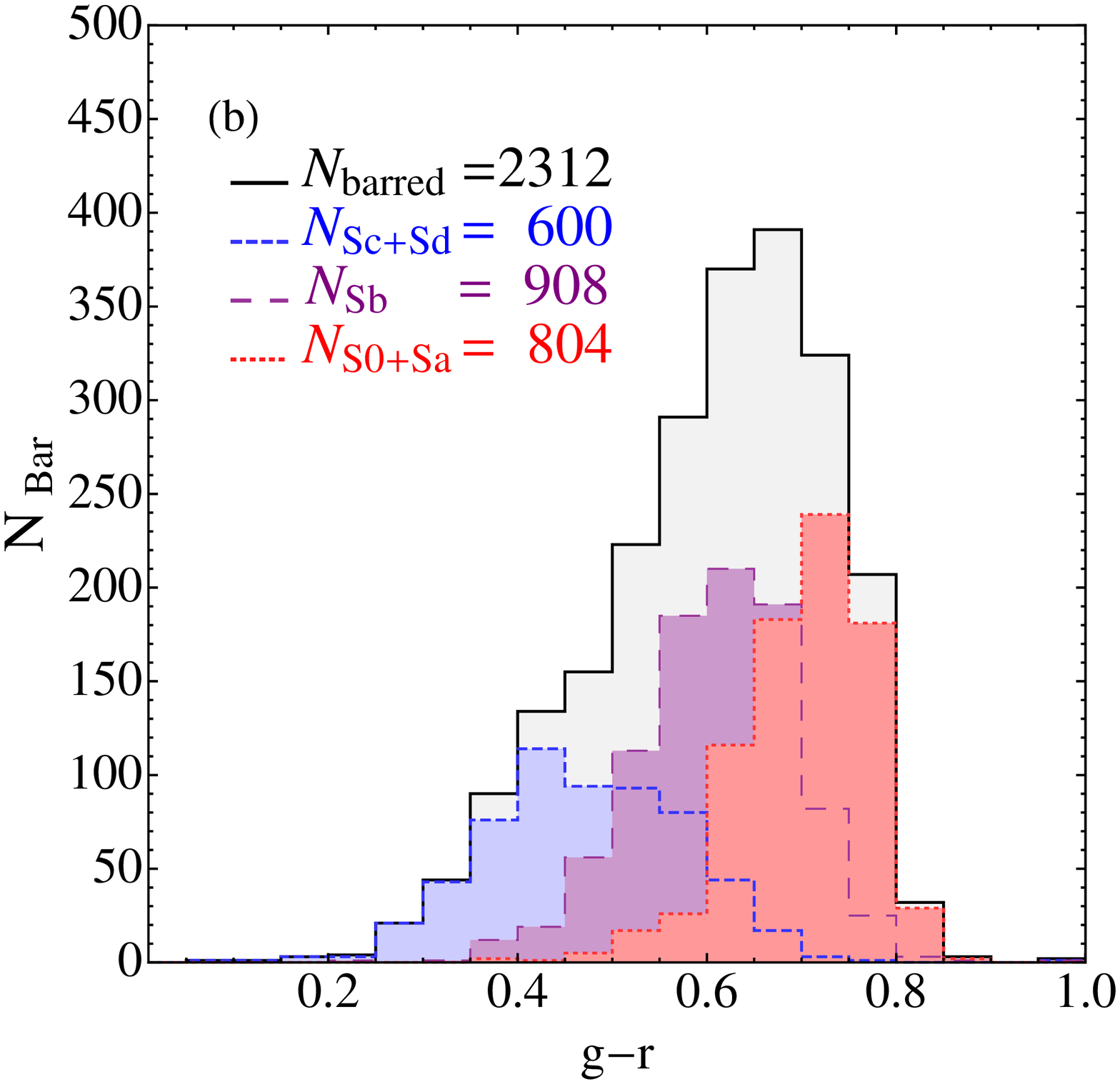}}}
\rotatebox{0}{\resizebox{4cm}{4cm}{\includegraphics{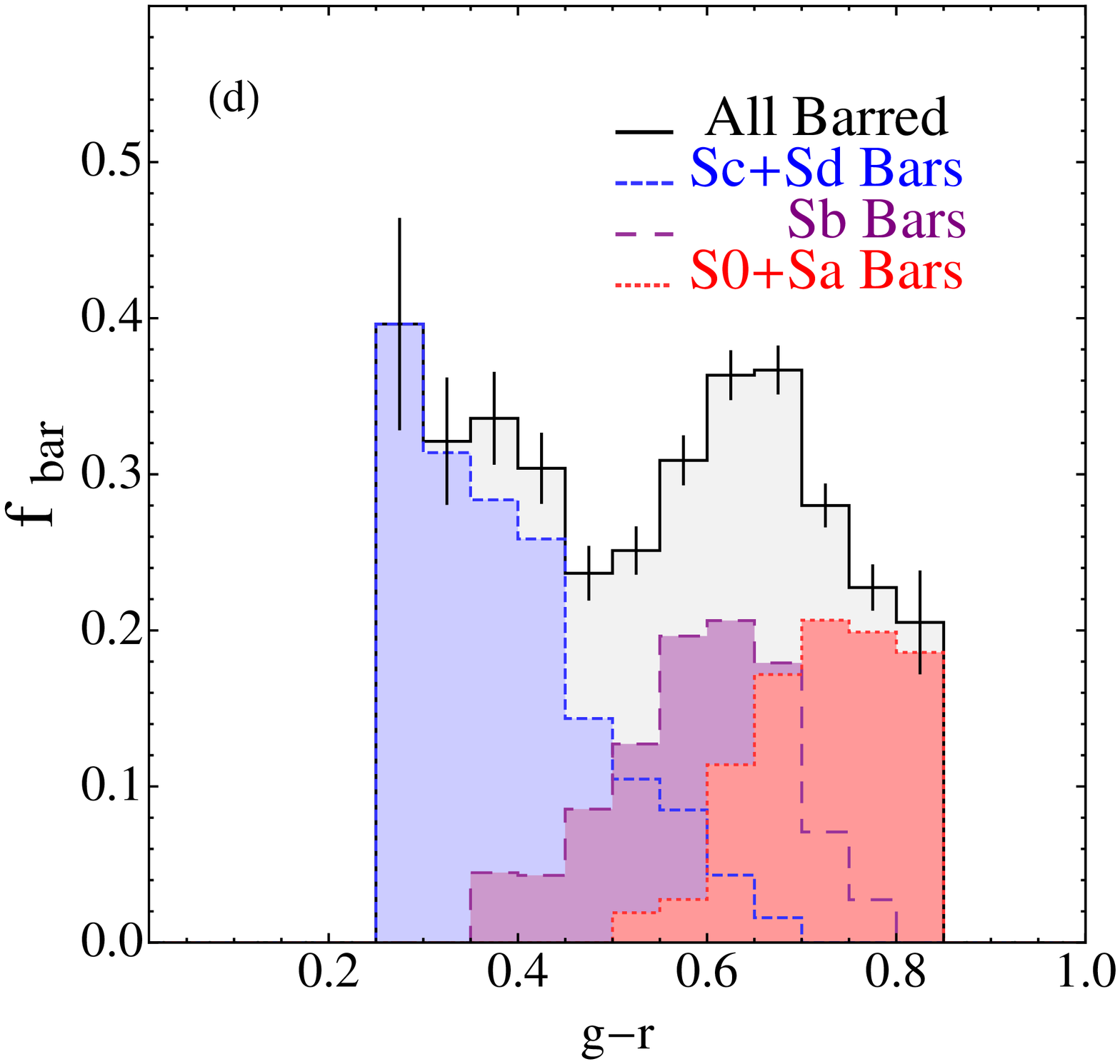}}}
\end{minipage}
\caption[Distributions of Bars, Rings and Lenses ]{\label{fig:BarDistributions1} {\bf Top}: Histograms and fractional distribution of (a/c) stellar mass as defined by \cite{Kauffmann:2003p97} and (b/d) (g-r) color for barred galaxies. The distribution of bars in S0+Sa(red), Sb(purple) and Sc+Sd(blue) galaxies as well as all barred galaxies (black) are shown. We find the bar fraction falls steeply from low masses to intermediate masses, $M \thicksim 10.2$,  and rises slowly and plateaus thereafter. With (g-r) color, we Þnd the bar fraction decreases from bluer colors to intermediate colors, (g-r) $\thicksim$0.5, and rises slowly thereafter.}
\end{figure}

Figure~\ref{fig:BarDistributions1} shows the fraction of barred spiral galaxies as a function of mass and $(g-r)$ color for our sample of disk objects. The top row shows a histogram of the number of bars while the bottom row shows the fraction of barred systems in each mass bin. 
Error bars on the fractions have been computed assuming binomial statistics. For the sake of clarity, we have excluded bins with fewer than 10 barred objects. The distribution is keyed to galaxy type where S0+Sa galaxies are represented by the red curve, Sb galaxies by the purple curve and Sc+Sd galaxies by the blue curve. The figure clearly shows that the bar fraction of local galaxies is bimodal with respect to stellar mass and color, with a strong break at $\log(M/M_\odot)$$\thicksim$10.2. Figure~\ref{fig:BarDistributions1}(c) shows that the bar fraction gradually decreases from $\thicksim$ 40\% for  low mass galaxies ($\log(M/M_\odot)$$\thicksim$9)  to $\thicksim 24\%$  for intermediate mass galaxies ($\log(M/M_\odot)$$\thicksim$10.2). It then gradually increases and plateaus at 30\% before truncating at  $\log(M/M_\odot)$$\thicksim$11.4. This break in mass mirrors the bimodality of bar fraction with T-Type as seen by \cite{Odewahn:1996p30} and which can also be seen in Figure~\ref{fig:BarDistributions1}(c) where late type galaxies clearly prefer the low mass peak, while early type galaxies prefer the high mass peak. The break in the fraction of barred spiral galaxies occurs at the characteristic mass where bimodality in galaxy properties has been observed \citep{Kauffmann:2003p7199,Shen:2003p70} and at which the blue cloud and red sequence becomes defined \citep{Baldry:2004p43}. Assuming that this is not a coincidence, and that a link exists between the decline in the bar fraction and the formation of blue and red sequences, one might expect that the distribution of color for barred galaxies would also be bimodal.
This is confirmed by Figure~\ref{fig:BarDistributions1}(d). The bar fraction initially decreases as the galaxies become redder from 40\% at (g-r)$\thicksim$0.22 to 24\% at (g-r)$\thicksim$0.42. Above (g-r)$\thicksim$0.42 the bar-fraction transitions and starts to increase as the galaxies become redder. The bar fraction starts to decrease again past  (g-r)$\thicksim$0.7. BJ08 find similar results for the bluest bins, while SE08 find similar results for bar fractions in the reddest bins.

The results just described gives us some insight into the importance of potential systematic effects which may be relevant for understanding results from other surveys. For example, a decreasing bar fraction from low to intermediate masses is also seen by BJ08. However BJ08 finds the bar fraction continues to decrease with increasing mass from $\log(M/M_\odot)\thicksim$10  to $\log(M/M_\odot)\thicksim$10.7, a range over which BJ08's data becomes sparse, while our sample remains abundant ($>$100 barred objects per bin). Since our own data shows a slightly increasing/constant bar fraction over this mass range, we conclude that the apparent discrepancy in the results of SE08 and BJ08 in this mass regime is probably due to the fraction of barred spiral galaxies being a strong function of stellar mass, with these surveys preferentially sampling below and above the characteristic mass of $\log(M/M_\odot)$$\thicksim10.2$ at which the bar fraction is at its minimum.

Our result also provides a possible solution to the discrepancy between the redshift evolution found by \cite{Sheth:2008p1183} and the works by \cite{Jogee:2004p16823} and \cite{Elmegreen:2004p1773}. SE08 finds that the bar fraction strongly decreases as a function of redshift. 
Their sample probes $M_{v}$$<$-21.7 at z$\thicksim$0.9 and $\log(M/M_\odot)$$>$10.2 whereas \cite{Jogee:2004p16823} find a near constant bar fraction using fainter samples with $M_{v}$$<$-19.3 and $M_{v}$$<$-20.6. Most likely, \cite{Jogee:2004p16823} sample further down the mass function, averaging over both the low mass and high mass peak whereas \cite{Sheth:2008p1183} are restricted to the high mass peak ($\log(M/M_\odot$)$>$10.2). Thus, the redshift evolution of bar fractions needs to be studied carefully, both above and below the transition mass.

\section{Discussion}

Our central conclusion is that the fraction of barred spiral galaxies
is strongly dependent on the mass and star-formation history of galaxies. This conclusions
naturally leads us to wonder whether our observations can be used to constrain scenarios for bar growth and/or destruction.
As has already been noted,
a number of galaxy properties like color \citep{Strateva:2001p18683,Baldry:2004p43}, luminosity \citep{Balogh:2004p47}, mass, surface mass density \citep{Kauffmann:2003p7199}, size \citep{Shen:2003p70} and concentration \citep{Shen:2003p70}
 exhibit bimodal characteristics. The fact that bimodality is manifested in so many parameters is perhaps not surprising given the strong internal correlations between them. To this list of bimodal properties an additional morphological signature, namely the bar fraction, can now be added. A minimum in the fraction of barred spiral galaxies occurring at the same mass at which bimodality manifests itself in stellar populations suggests  that whatever process is causing the creation of the red and blue sequence is either influencing, or being influenced by, structural changes in the galaxies in which bars become rarer. 

\begin{figure}[t!]
\unitlength1cm
\begin{minipage}[t]{4.0cm}
\rotatebox{0}{\resizebox{8cm}{11cm}{\includegraphics{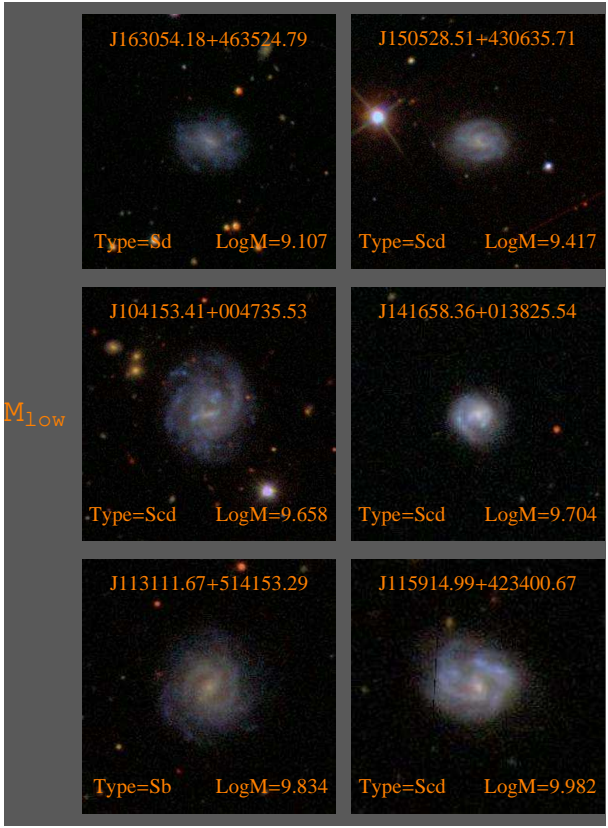}}}
\end{minipage}
\caption[Bimodality in Bars]{\label{fig:BimodalBarsMontage1} 
The two columns shows a random sample of bars below the transition mass of $10.2$. The J2000 object identifier is listed at the top, the type in the bottom left corner and the mass on the right. Objects are arranged in order of increasing mass.}
\end{figure}

\begin{figure}[t!]
\unitlength1cm
\begin{minipage}[t]{4.0cm}
\rotatebox{0}{\resizebox{8cm}{11cm}{\includegraphics{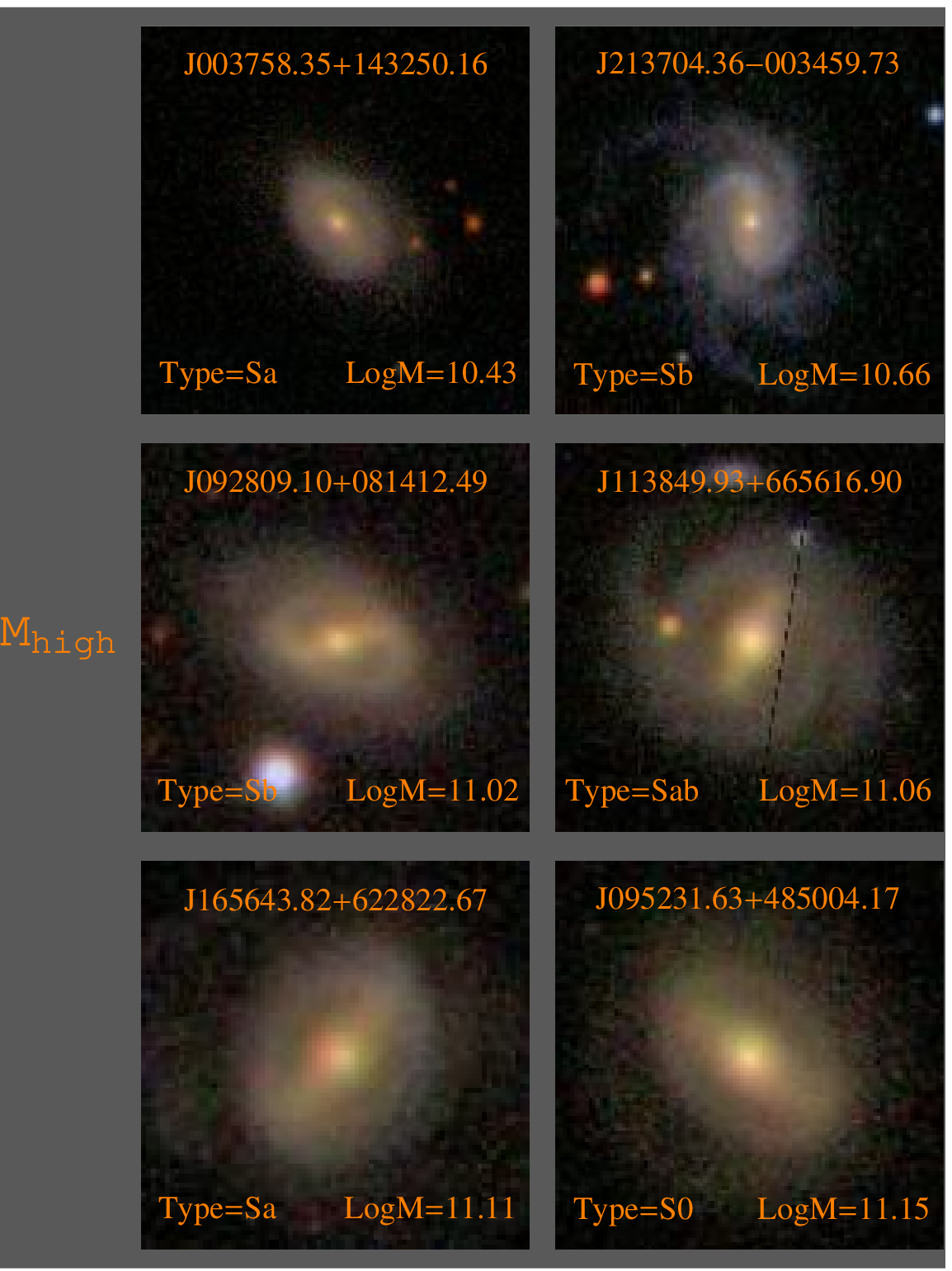}}}
\end{minipage}
\caption[Bimodality in Bars]{\label{fig:BimodalBarsMontage2} 
The two columns shows a random sample of bars above the transition mass of 10.2. The J2000 object identifier is listed at the top, the type in the bottom left corner and the mass on the right. Objects are arranged in order of increasing mass.}
\end{figure}

It has been shown by \cite{Elmegreen:1985p12037}
that the characteristics of bars themselves change along the Hubble Sequence. Bars in early type galaxies are longer, stronger, show a flatter light profile and a strong correlation with grand design 2-spiral arm structures compared to bars in late type galaxies which show an exponential light profile and more multiple armed or flocculent arm structure. This bimodality in bar type may be directly related to the mass bimodality in the fraction of barred spirals. To illustrate this, Figures~\ref{fig:BimodalBarsMontage1} and \ref{fig:BimodalBarsMontage2}  show representative examples of bars below and above the transition mass threshold of $\log(M/M_\odot) \thicksim 10.2$ sorted by increasing mass. The first noticeable difference between the two panels is the color of the galaxy, where the lower mass galaxies are blue while the higher mass galaxies are red, as expected. Galaxies in the higher mass bin have definite bulges while those in the lower mass bin have no bulge or a very tiny bulge and more flocculent arm structure. There is a possible indication of increasing bulge presence with mass in low-mass barred galaxies. If this is the case we would expect bar fraction to be keyed to central concentration, which is an easily measurable (albeit crude) proxy for bulge strength.

\begin{figure}[t!]
\unitlength1cm
\begin{minipage}[t]{4.0cm}
\rotatebox{0}{\resizebox{4cm}{4cm}{\includegraphics{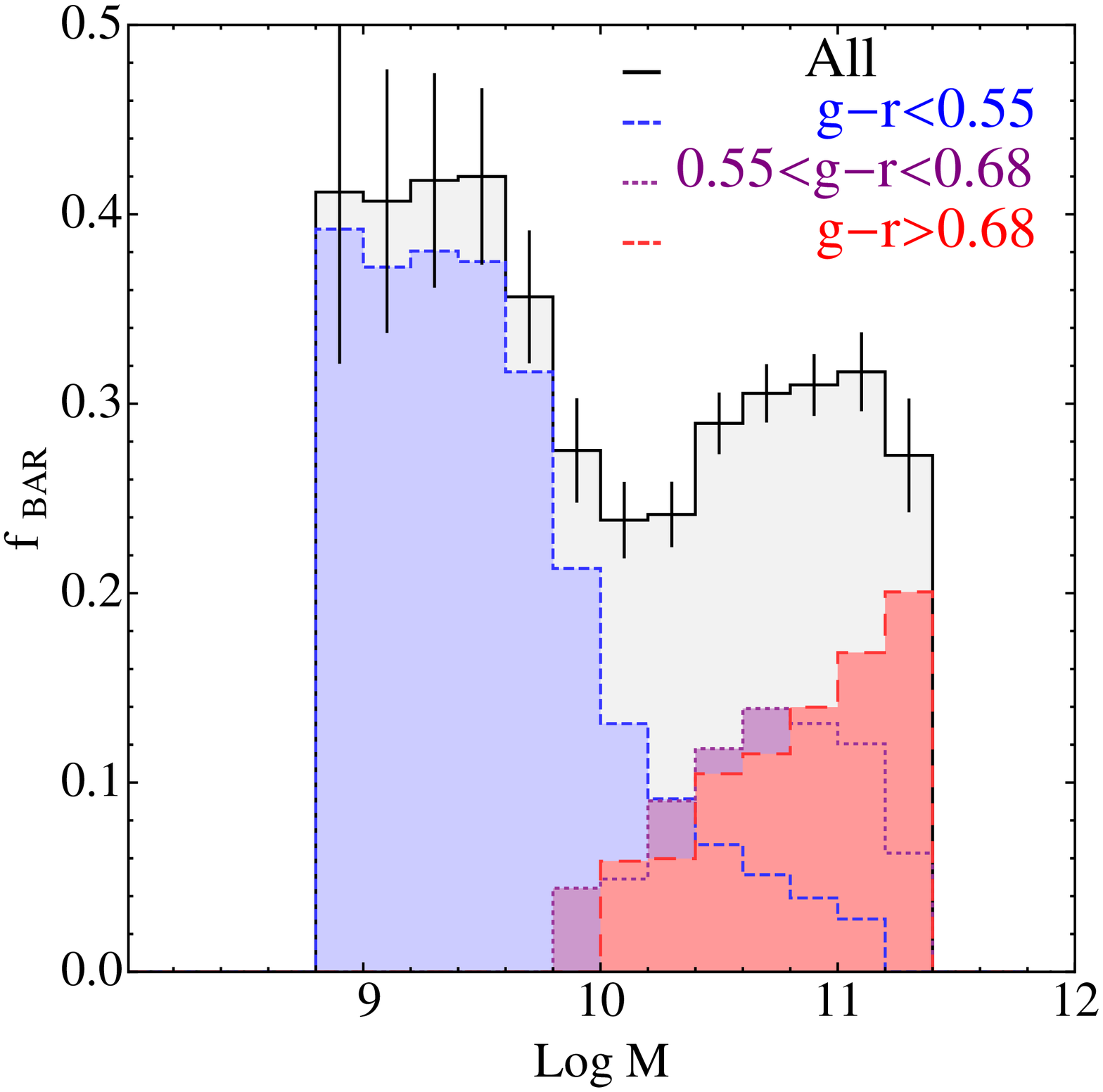}}}
\end{minipage}
\begin{minipage}[t]{4.0cm}
\rotatebox{0}{\resizebox{4cm}{4cm}{\includegraphics{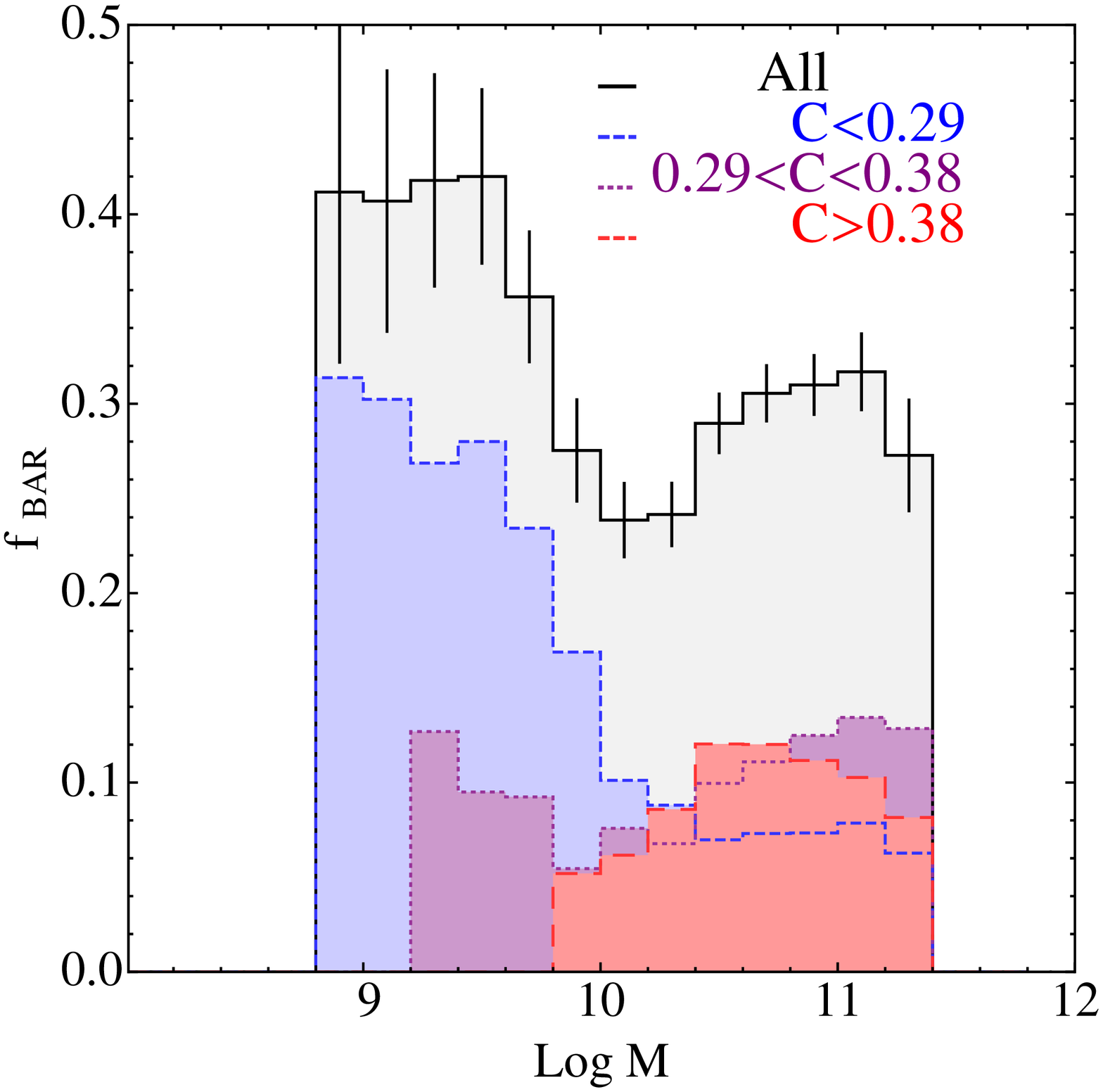}}}
\end{minipage}
\caption[Bar fraction vs mass keyed to color and concentration]{\label{fig:BarKeyedToCG}Bar fraction as a function of Mass keyed to color  (left) and concentration(right) in three quantile bins. Blue (small dash) indicates the lowest quantile, purple (dotted) the intermediate range and red (dashed) the highest quantile.}
\end{figure}

\begin{figure}[t!]
\unitlength1cm
\begin{minipage}[t]{4.0cm}
\rotatebox{0}{\resizebox{4cm}{4cm}{\includegraphics{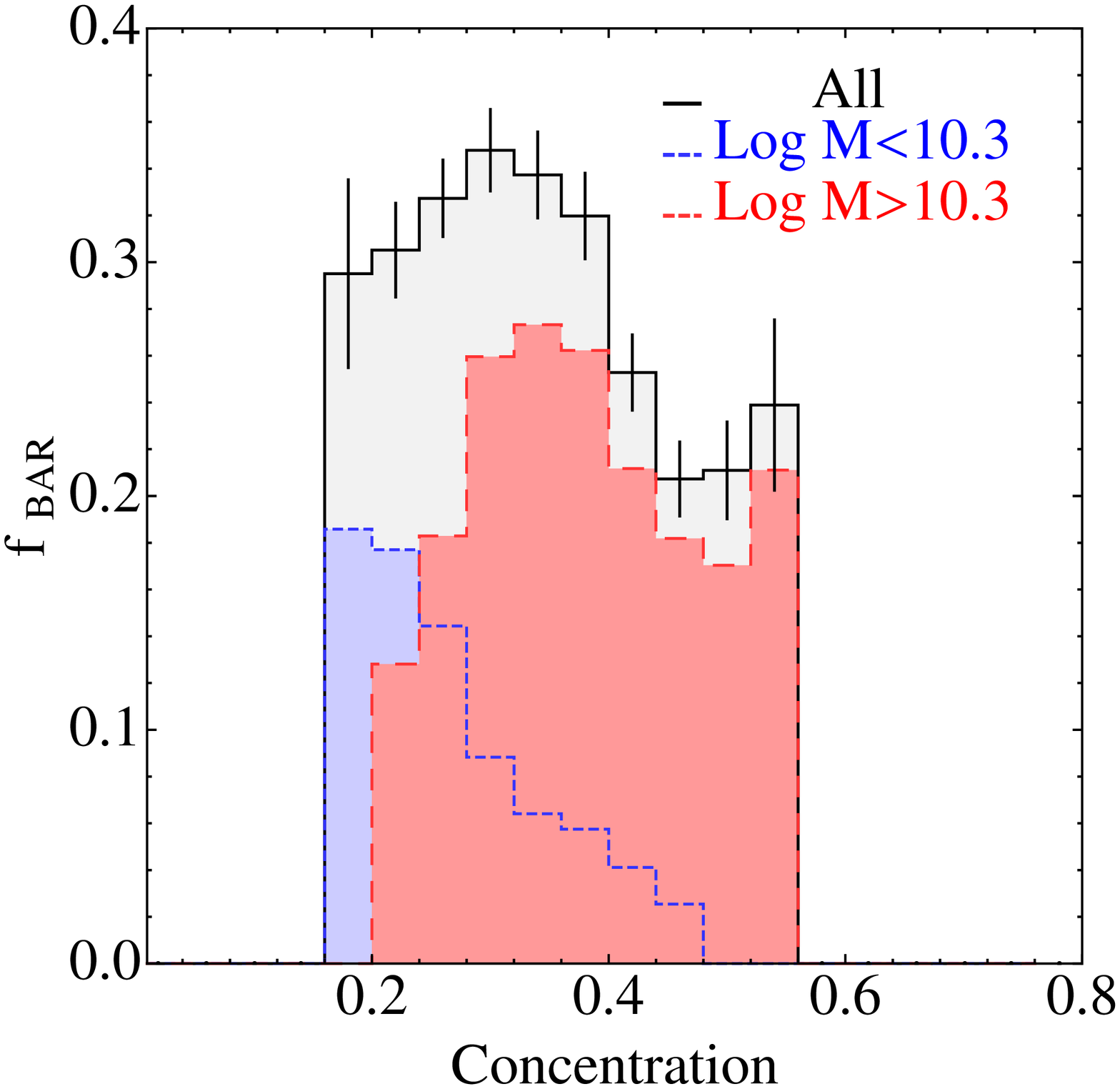}}}
\end{minipage}
\begin{minipage}[t]{4.0cm}
\rotatebox{0}{\resizebox{4cm}{4cm}{\includegraphics{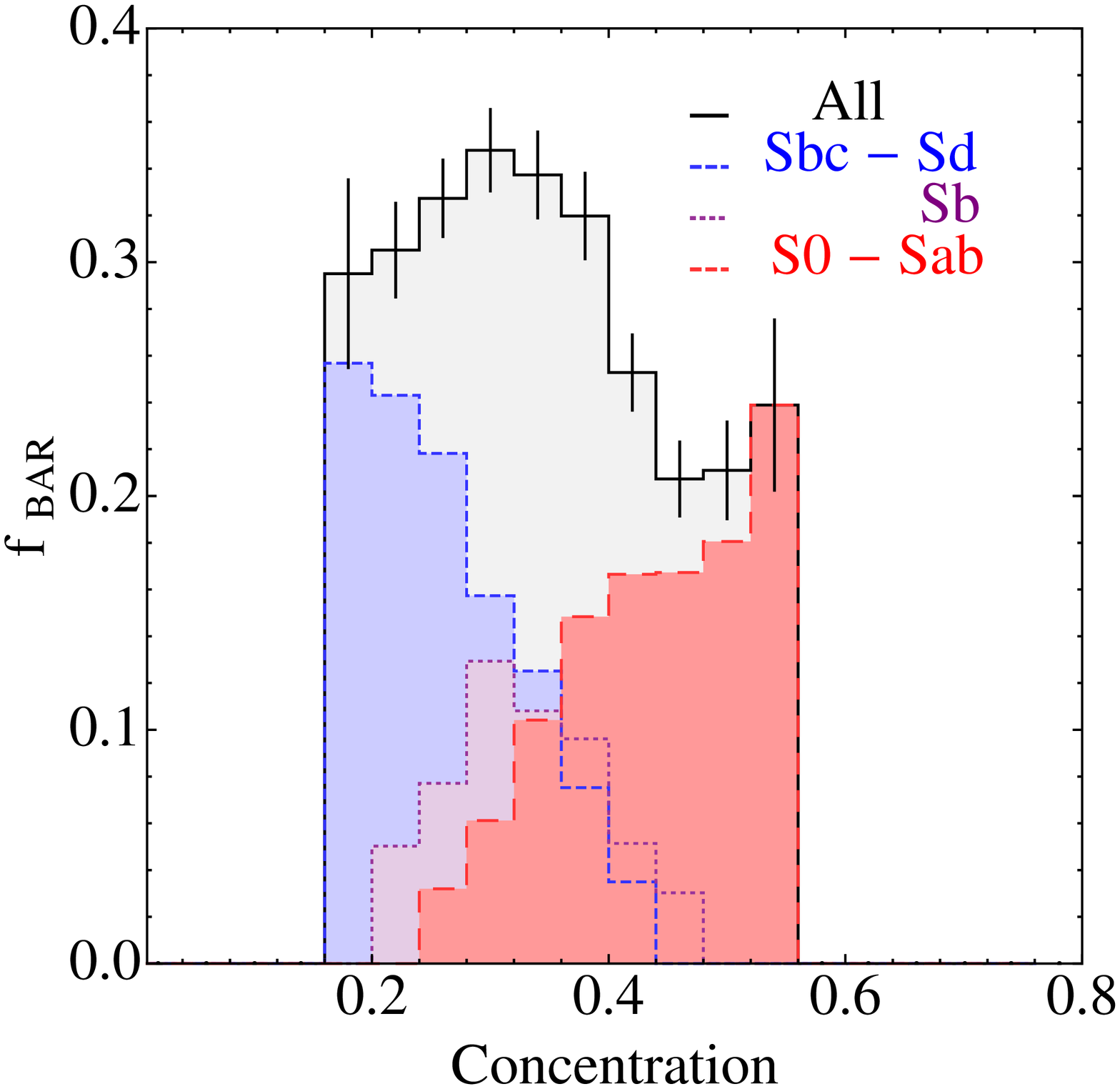}}}
\end{minipage}
\caption[Bar fraction vs conc keyed to mass]{\label{fig:BarKeyedToConcM}Bar fraction as a function of central concentration keyed to (left panel) stellar mass, where the blue (dotted) curve shows galaxies below the transition mass and red (dashed) curve shows galaxies above the transition mass, and (right panel) Hubble type.}
\end{figure}

To test this idea, Figure~\ref{fig:BarKeyedToCG} shows the fractional histograms relating the bar fraction to mass keyed to (g-r) color (left panel) and concentration\footnote{Defined as the ratio of flux within an inner and outer elliptical aperture
 determined from the sky-subtracted, intensity-weighted, second-order moment of the image. 
 The major and minor axes of the outer aperture are normalized so that the total area within the ellipse is the area of the galaxy. The inner aperture is defined by scaling these axes down by a factor of 3.} (right panel) for objects in our sample. We find blue objects occupy the low mass peak and redder objects occupy the high mass peak as expected. From the right panel of Figure~\ref{fig:BarKeyedToCG}, we find that the low mass peak is dominated by low concentration galaxies but low concentration galaxies span the whole range in mass. What is more interesting is that for objects with $\log(M/M_{\odot})$$<$10.2 (low mass peak), at a given stellar mass the bar fraction decreases as concentration increases, while in the high mass peak ($\log(M/M_{\odot})>$10.2) the reverse is true: as concentration increases, the bar fraction increases. (It should be noted that the bar fractions are roughly the same in the two highest concentration bins although there is a slight mass dependence). This can be seen more clearly in Figure~\ref{fig:BarKeyedToConcM} which shows bar fraction versus concentration keyed to (left) two mass bins and (right) galaxy type. For low mass galaxies, bar fraction decreases with concentration whereas for high mass galaxies, the bar fraction dependence is more complicated. These trends are clearer with Hubble type. Bar fractions in Sbc and later galaxies clearly decrease with increasing central concentrations whereas they increase for Sab and earlier types. Sb galaxies appear to be a bridge between the two populations. 

The different ways in which bar fraction varies with central concentration (and type) above and below $\log(M/M_{\odot})$$\thicksim$10.2 suggests that bar formation (or destruction) may be operating in fundamentally different ways in these two mass regimes. Do numerical simulations shed any light on these conclusions?

\subsection{Bar destruction mechanisms in the low mass peak}
The susceptibility of bars to destruction by  a central mass concentration (CMC) has been studied in many simulations \citep{Norman:1996p8611,Athanassoula:2002p7991,Shen:2004p11908,Hozumi:1999p11909,Bournaud:2005p11901,Athanassoula:2005p7976,Curir:2008p12620}. Most simulations agree that it is possible to destroy or severely reduce a bar in different circumstances, though the results on the actual mass of the central concentration required to destroy a bar varies, and is usually much higher than $\log(M/M_\odot) \thicksim 10.3$. The CMC can be due to stellar mass but also due to the build up of gas and dust. It has been shown that barred galaxies have a larger concentration of CO \citep{Sheth:2005p7966,Sakamoto:1999p7925} and PAH emission \citep{Regan:2006p12563} than unbarred galaxies. Some simulations suggest that while the growth of the CMC may not completely dissolve a bar \citep{Shen:2004p11908} the gas flow to the center aided by the bar is itself responsible for bar destruction \citep{Friedli:1993p8615}. \cite{Bournaud:2005p11901} suggests the transfer of angular momentum between the in-falling gas and the bar can severely weaken the bar. These two processes, the build up of a central mass concentration and the transfer of angular momentum from infalling gas to bars in stellar mass dominated systems could account for the decrease in bar fraction we observe in low mass galaxies. \cite{Athanassoula:2005p7976} find that `massive disk' (or late type) galaxies are very prone to bar dissolution where even a 5\% central mass concentration can destroy a bar and a 1\% mass concentration can considerably weaken exponential bars.
\cite{Curir:2008p11957} find bars are more easily destroyed in their simulations of $\log(M/M_\odot) \thicksim 10.3$ galaxies than in their lower mass ($\log(M/M_\odot) \thicksim 9.7$), dark matter dominated galaxies, which agrees with our results.

Another possible explanation for our results might be found in the local environment of galaxies, which seems particularly relevant to understanding bars in low-mass/low-concentration galaxies. The halo and gas play a very important role in such systems where they tend to dominate the dynamics. Numerical simulations show that low mass, gas dominated galaxies are very prone to instabilities and bar formation. This could explain why the bar fraction is higher in the low mass, blue peak as opposed to the high mass, red peak. A possible trigger could be minor satellite impacts (10:1 ratio) which have been shown to cause bar instabilities in axisymmetric disks \citep{Dubinski:2008p12592}. In addition asymmetries in the dark matter distribution or internal instabilities have also been considered as mechanisms for triggering bar instabilities \citep{Curir:2008p11957}. As the galaxy converts more of its gas to stars, it becomes harder for bars to be triggered by instabilities, thus naturally explaining the trend of lower bar fraction with increasing stellar mass in the low mass peak. 

\subsection{Bar formation mechanisms in the high mass peak}
At the high-mass end a number of numerical simulations suggest bars in galaxies with massive halos remain stable once formed \citep{Athanassoula:2005p7976}. In the low redshift Universe, we find a near constant bar fraction at masses greater than 10.4 (log units), as does SE08. However, SE08 also finds that this slope evolves with redshift where the highest mass objects have the highest bar fraction at $z\thicksim1$ and intermediate mass objects build up their bar fraction between $0<z<1$. Thus it appears the bar formation mechanism in the high mass peak is either more efficient or more stable for higher mass galaxies than for the intermediate mass galaxies (in contrast to the low mass peak). If we assume that this is true then perhaps bars forming in high redshift galaxies with intermediate masses ($10.0<\log(M/M_\odot)<10.5$) do not have halos with sufficient mass to prevent destruction by processes such as merging or gas inflow. These galaxies have not yet had the opportunity to build a stabilizing halo through a steady diet of low-mass mergers or through infall into a larger group. In a companion paper (in prep), we will study the effect of environment and AGN on bar formation/destruction. 

\subsection{Evolution of bar fraction}
Identifying the causes of bar destruction will be very important to understanding the redshift evolution of barred galaxies and possibly the formation of the red and blue sequences. This may be testable in a fairly straightforward way, if we note that there appears to be two methods for destroying bars in low mass galaxies. The first is related to increasing galaxy mass and the second to increasing central concentration. Numerical simulations have shown that exponential bars common in low mass galaxies are prone to bar destruction as the central concentration increases while flat surface density strong bars which occupy high mass galaxies are not destroyed by a central mass concentration, but may have their bars shortened \citep{Athanassoula:2005p7976}. This suggests that an analysis of bar sizes as a function of cosmic epoch, Hubble type and environment might prove interesting. 

\section{Conclusions}

We find that the fraction of barred spiral galaxies is a strongly bimodal
function of stellar mass, with a minimum near $\log(M/M_\odot)\thicksim$10.2. This is also
the characteristic mass at which bimodality is seen in the stellar populations of galaxies.
This suggests  that whatever process is causing the creation of the red and blue sequences is 
linked in some way to the formation (or destruction) of bars. Because estimates
of the local  barred galaxy fraction depend sensitively on the stellar mass range being probed,
our results suggest that inconsistencies in the reported fractions of barred spiral galaxies can
be understood rather simply as a selection effect in which surveys have
obtained different results because they have been probing galaxies in different stellar mass ranges.

\acknowledgments
\noindent{\em Acknowledgments}

We would like to thank the anonymous referee for many useful comments and suggestions.

Funding for the SDSS and SDSS-II has been provided by the Alfred P. Sloan Foundation, the Participating Institutions, the National Science Foundation, the U.S. Department of Energy, the National Aeronautics and Space Administration, the Japanese Monbukagakusho, the Max Planck Society, and the Higher Education Funding Council for England. The SDSS Web Site is http://www.sdss.org/.

\end{document}